\documentclass[12pt]{article}
\usepackage[cp1251]{inputenc}
\usepackage[russian]{babel}
\usepackage{amsmath}
\usepackage{amssymb}
\usepackage{amsfonts}
\usepackage{cite}
\usepackage{cmap}
\usepackage[pdftex,unicode]{hyperref}

\newcommand{\itemaa}{\smallskip${{\bigstar}}$\hspace{0.1cm}}
\newcounter{inum}
 
\newtheorem{lemma}{Лемма}
\newtheorem{theorem}{Теорема}

\newtheorem{hyp}{Гипотеза}

\numberwithin{equation}{section}

\newcounter{meq}
\def\mmeq#1{\refstepcounter{meq}\begin{equation*} #1\eqno{(\hbox{S}\themeq)}\end{equation*}}

\def\seqref#1{(S\ref{#1})}

\def\eqa#1{\begin{equation}\begin{aligned}#1\end{aligned}\end{equation}}
\def\eq#1{\begin{equation}#1\end{equation}}

\def\seq#1{\begin{equation*}#1\end{equation*}}
\def\seqs#1{\begin{equation*}\begin{array}{c}#1\end{array}\end{equation*}}
\def\pdfrac#1#2{\frac{\partial #1}{\partial #2}}

\def\phi{\varphi}

\def\be{\begin{equation}}
\def\ee{\end{equation}}
\def\ba{\begin{aligned}} 
\def\ea{\end{aligned}}

\newcounter{theo}

\newcounter{lem}

\newcounter{prop}
\newcounter{rem}

\newcounter{defi}

\newcounter{examp}

\def\pfrac#1#2{\frac{\partial #1}{\partial #2}}

\numberwithin{equation}{section}
\title{Классификация полудискретных уравнений гиперболического типа. Случай симметрий пятого порядка.}

\author{Р.Н. Гарифуллин.\thanks{\rm Исследование выполнено за счет гранта Российского научного фонда № 21-11-
00006, https://rscf.ru/project/21-11-00006/.}
\thanks{\rm Институт математики с вычислительным центром Уфимского федерального исследовательского центра РАН, Уфа, Россия.}}
\begin{document}

\maketitle {\small
\begin{quote}
\noindent{\bf Аннотация. } В работе проводится квалификация полудискретных уравнений гиперболического типа. Исследуется класс уравнений вида
\seq{\frac{du_{n+1}}{dx}=f\left(\frac{du_{n}}{dx},u_{n+1},u_{n}\right),} здесь неизвестная функция $u_n(x)$ зависит от одной дискретной $n$ и одной непрерывной $x$ переменных. 
Квалификации основывается на требовании существования высших симметрий. Рассматривается случай когда симметрия имеет порядок 5 в непрерывном направлениях. В результате получен список 4 уравнений с требуемыми условиями.
Для одного из найденых уравнений построено представление Лакса.
\medskip

\noindent{\bf Ключевые слова:}  {интегрируемость, высшая симметрия, классификация, полудискретное уравнение, гиперболический тип.
}
\medskip
\end{quote}
 }

Эта работа является продолжением статьи \cite{g23} в которой проведена классфикация уравнений с симметриями третьего порядка. Исследуются  полудискретные уравнения гиперболического типа\eq{u_{n+1,x}=f(u_{n,x},u_{n+1},  u_{n},x )\label{seq},} где неизвестная функция $u_n(x)$ зависит от одной дискретной $n$ и одной непрерывной $x$ переменных. Здесь и ниже используется обозначение $$u_{k,x}=\frac{du_{k}}{dx},\ u_{k,xx}=\frac{d^2u_{k}}{dx^2},\ u_{k,xxx}=\frac{d^3u_{k}}{dx^3},\ u_{k,t}=\frac{du_{k}}{dt},\ u_{k,\tau}=\frac{du_{k}}{d\tau}.$$

Наиболее известным представителем этого класса является одевающая цепочка, подробное исследование которой проведено в статье А.П.~Веселова и А.Б.~Шабата  \cite{vs93}\eq{u_{n+1,x}+u_{n,x}=u_{n+1}^2-u_{n}^2\label{dr},} которая возникла как преобразование Бэклунда для модифицированного уравнения Кортевега де Вриза:
\eq{u_{n,t}=u_{n,xxx}-6u_n^2u_{n,x}\label{kdv}.} С другой стороны уравнение \eqref{kdv} можно рассматривать как высшую симметрию уравнения \eqref{dr}. По дискретному направлению высшая симметрия уравнения \eqref{dr} имеет вид \eq{u_{n,\tau}=\frac{1}{u_{n+1}+u_{n}}-\frac{1}{u_{n}+u_{n-1}}\label{vol}} и является известным дифференциально-разностным уравнением \cite{y83,y06}. В статье Р.И.~Ямилова  \cite{y90} был приведен ряд примеров троек уравнений типа \eqref{dr}-\eqref{vol}.

В недавней работе \cite{gh21} был предложен метод построения высших симметрий уравнений вида \eqref{seq}. Было показано, что высшая симметрия в непрерывном направлении является эволюционным уравнением вида:
\eq{u_{n,t}=\frac{d^Nu_n}{dt^N}+F\left(x,u_n, \, \frac{du_n}{dt}, \dots, \frac{d^{N-1}u_n}{dt^{N-1}}\right).      \label{scalar}
}Такие уравнения называются уравнениями с постоянной сепарантой \cite{ms12}. С другой стороны полудискретного уравнения вида \eqref{seq} совместностное с уравнением вида \eqref{scalar} можно рассматривать как автопреобреобразование Бэклунда. Поэтому уравнения \eqref{scalar} сами по себе также являются инегрируемыми уравнениями. Список таких уравнений порядков 3 и 5 был приведен в обзоре \cite{ms12}, в нем также подробно изложена история вопроса.

В этой работе мы будет рассматривать уравнения пятого порядка и для них искать полудискретные уравнения вида \eqref{seq}. Список таких уравнений пятого порядка имеет вид:
\begin{align}
&u_t=u_{5}+5 u u_{3}+5 u_1 u_2+5 u^2 u_1, \label{tst1} \\
\label{tst2}
&u_t=u_{5}+5 u u_{3}+{25 \over 2} u_1 u_2+5 u^2 u_1,  \\
\label{tst3}
&u_t=u_{5}+5 u_1 u_{3}+{5 \over 3} u_1^3,  \\ 
\label{tst4}
&u_t=u_{5}+5 u_1 u_{3}+{15\over 4}u_2^2 + {5 \over 3} u_1^3, \\
\label{tst5}
&u_t=u_{5}+5 (u_1-u^2) u_{3}+5 u_2^2-20 u u_1 u_2-5 u_1^3+5 u^4 u_1, \\
\label{tst6}
&u_t=u_{5}+5 (u_2-u_1^2) u_{3}-5 u_1 u_2^2+u_1^5, \\
\label{tst7}
&\begin{aligned}
u_t&=u_{5}+5 (u_2-u_1^2-\lambda_1^2 e^{2u}-\lambda_2^2 e^{-4u}) \, u_{3}-5 u_1 u_2^2-15 (\lambda_1^2 e^{2u}-4 \lambda_2^2 e^{-4u})\, u_1 u_2 \\[1mm]
&+u_1^5-90 \lambda_2^2 e^{-4u}\, u_1^3 +5(\lambda_1^2 e^{2u}+\lambda_2^2 e^{-4u})^2\, u_1,
\end{aligned}\\[2mm]
\label{tst8}
&\begin{aligned}
u_t&=u_{5}+5 (u_2-u_1^2-\lambda_1^2 e^{2u}+\lambda_2 e^{-u}) \, u_{3}-5 u_1 u_2^2-15 \lambda_1^2 e^{2u} \, u_1 u_2 \\[1mm]
&+u_1^5+5(\lambda_1^2 e^{2u}-\lambda_2 e^{-u})^2 \, u_1, \quad \lambda_2\ne 0, 
\end{aligned}\\[2mm]
\label{tst9}
&\begin{aligned}
u_t&=u_{5}-5\frac{u_2 u_{4} }{ u_1}+ 5\frac{u_2^2 u_{3}}{ u_1^2}+5\left(\frac{ \mu_1}{ u_1}+\mu_2 u_1^2\right)u_{3}-5\left(\frac{ \mu_1}{ u_1^2}+\mu_2 u_1\right)u_2^2
\\[1mm]
&-5 \frac{\mu_1^2}{ u_1}+ 5 \mu_1\mu_2 u_1^2 +\mu_2^2 u_1^5,
\end{aligned}\\
\label{tst10}
&\begin{aligned}
u_t&=u_{5}-5\frac{u_2 u_{4}}{u_1}-\frac{15}{4 }\,\frac{u_{3}^2}{u_1}+\frac{ 65}{4}\,\frac{u_2^2 u_{3}}{ u_1^2} +5\left(\frac{\mu_1}{u_1}+\mu_2 u_1^2\right)\, u_{3}
-\frac{135}{16 }\frac{u_2^4}{u_1^3} 
\\
&-5\left(\frac{7 \mu_1}{4 u_1^2}-\frac{\mu_2 u_1}{2}\right) u_2^2-5 \frac{\mu_1^2}{u_1}+ 5 \mu_1\mu_2 u_1^2 +\mu_2^2 u_1^5,
\end{aligned}
\end{align}
\begin{align}
\label{tst11}
&\begin{aligned}
u_t&= u_{5}-\frac{ 5 }{2}\,\frac{ u_2u_{4} }{ u_1}-\frac{5}{4}\,\frac{u_{3}^2}{ u_1}+5\frac{u_2^2 u_{3}}{u_1^2} +\frac{5\, u_2 u_{3}}{2 \sqrt{u_1}}
-5 (u_1-2 \mu u_1^{1/2}+\mu^2) \, u_{3} -\frac{ 35}{16}\,\frac{u_2^4}{ u_1^3}  \\
 & -\frac{5}{3}\,\frac{u_2^3}{u_1^{3/2}} +5\Big (\frac{ 3 \mu^2}{ 4 u_1} -\frac{\mu }{\sqrt{u_1}}+\frac{1}{4}\Big)\, u_2^2
+ \frac{ 5}{ 3}\, u_1^3  
 - 8 \mu u_1^{5/2}+15 \mu^2 u_1^2-\frac{40}{3}\,\mu^3 u_1^{3/2},
\end{aligned}
\end{align}
\begin{align}
&\begin{aligned}\label{tst12m}
u_t&=u_{{{ 5}}}+\frac52\,{\frac { f-u_{1}}{{f}^{2}}}\,u_{{2}}u_{{{ 4}}}+\frac54\,{\frac {2\,f -u_{1}}{{f}^{2}}}\,u_3^{2} +5\,\mu\, ( u_{1}+f ) ^{2}u_{{{3}}}\\
&+\frac54\,{\frac {4\,{u_{1}}^{2}-8\,u_{1}f+{f}^{2}}{{f}^{4}}}\,u_2^{2}u_{{{ 3}}} +{\frac {5}{16}}\,{\frac {2-9\,u_{1}^{3}+18\,u_{1}^{2}f}{{f}^{6}}}\,u_2^{4}\\
&+\frac {5\mu}{4}\,{\frac {( 4\,f-3\,u_{1} )( u_{1}+f )^{2}}{{f}^{2}}}\,u_2^{2}+{\mu}^{2} ( u_{1}+f ) ^{2}\big( 2\,f ( u_{1}+f ) ^{2}-1\big),
\end{aligned}
\end{align}
\begin{align}
&\begin{aligned}\label{tst13m}
u_t&=u_{{{ 5}}}+\frac52\,\frac {f- u_{1}}{f^{2}}\, u_2u_4+\frac54\,\frac {2\,f- u_1}{f^2}\, u_3^{2}-5\,\omega\, ( {f}^{2}+u_1^{2} ) u_3 \\
&+\frac54\,{\frac {4\,u_1^{2}-8\,u_{1}f+{f}^{2}}{{f}^{4}}}\,u_2^{2}u_3+{\frac {5}{16}}\,\frac {2-9\,u_1^{3}+18\,u_1^{2}f}{f^6}\,u_2^{4}
\\
&+\frac54\,\omega\,\frac {5\,u_1^{3} -2\,u_1^{2}f-11\,u_{1}{f}^{2}-2}{f^2}\, u_2^{2}-\frac52\,{\omega'}\, ( u_1^{2}-2\,u_{1}f+5\,{f}^{2} )u_{1}u_2 \\
&+5\,{\omega}^{2}u_{1}{f}^{2} ( 3\,u_{1}+f ) ( f-u_{1}),
\end{aligned}
\end{align}
\begin{align}
&\begin{aligned}\label{tst14m}
u_{{t}}&=u_5+\frac52\,{\frac {f-u_{1}}{{f}^{2}}}\,u_2u_4+\frac54\,{\frac {2\,f -u_{1}}{{f}^{2}}}\,u_3^{2}
+\frac54\,{\frac {4\,u_1^{2}-8\,u_{1}f+{f}^{2}}{{f}^{4}}}\,u_2^{2}u_3
\\
&+{\frac {5}{16}}\,{\frac {2-9\,{u_{1}}^{3}+18\,{u_{1}}^{2}f}{{f}^{6}}}\,u_2^{4}+5\,\omega\,{\frac {2\,{u_{1}}^{3}+{u_{1}}^{2}f-2\,u_{1}{f}^{2}+1}{{f}^{2}}}\,u_2^{2}
\\
& -10\,\omega\,u_{{{3}}} ( 3\,u_{1}f+2\,u_1^{2}+2\,{f}^{2} )-10\,{\omega'} ( 2\,{f}^{2}+u_{1}f+{u_{1}}^{2} )\,u_{1}u_{{2}}\\
& +20\,{\omega}^{2}u_{1} ( {u_{1}}^{3}-1 )( u_{1}+2\,f ),
\end{aligned}
\end{align}
\begin{align}
\label{tst15m}
&\begin{aligned}
u_t&=u_5+\frac52\,\frac {f-u_1}{f^2}\,u_2u_4+\frac54\,{\frac {2\,f-u_{1}}{{f}^{2}}}\,u_3^{2}-5\,c\frac {{f}^2+u_1^2}{{\omega}^{2}}\,u_3
\\
&+\frac54\,{\frac {4\,u_1^{2}-8\,u_{1}f+{f}^{2}}{{f}^{4}}}\,u_2^{2}u_3 +\frac {5}{16}\,\frac {2-9\,{u_{1}}^{3}+18\,u_1^{2}f }{{f}^{6}}\,u_2^{4}
\\
&-10\,\omega\, ( 3\,u_{1}f+2\,u_1^{2}+2\,{f}^{2} )\,u_3-\frac54\,c\,\frac { 11\,u_{1}{f}^{2}+2\,u_1^{2}f +2-5\,u_1^{3} }{{\omega}^{2}{f}^2}\,u_2^{2}
\\
&+5\,\omega\,\frac {2\,u_1^{3}+u_1^{2}f-2\,u_{1}{f}^{2}+1}{f^2}\,u_2^{2}+5\,c\,{\omega'}\,\frac {u_1^{2}+5\,{f}^{2}-2\,u_{1}f }{{\omega}^3}\,u_{1}u_2
\\
&-10\,{\omega'}\,(2\,{f}^{2}+u_{1}f+{u_{1}}^{2})\,u_1u_2 +20\,{\omega}^{2}u_{1} ( u_1^{3}-1 )  ( u_{1}+2\,f ) 
\\
&+40\,{\frac {c\,u_{1}{f}^{3} ( 2\,u_{1}+f ) }{\omega}}+5\,{\frac {{c}^{2}u_{1}{f}^{2} ( 3\,u_{1}+f )( f-u_{1} ) }{{\omega}^{4}}},\ \ c\ne0.
\end{aligned}
\end{align}
{\it Здесь} $\lambda_1, \lambda_2,\mu, \mu_1, \mu_2$ {\it и} $c$ --- {\it параметры},
{\it функция}  $f(u_1)$ {\it является решением алгебраического уравнения}
\begin{equation}\label{alg1}
(f+u_1)^2(2f-u_1)+1=0,
\end{equation}
{\it а} $\omega(u)$ --- {\it это любое непостоянное решение дифференциального уравнения}
\begin{equation}\label{Waier}
\omega '^2=4\, \omega^3+c. \qquad \square
\end{equation}

Приведенный список уравнений отличается от списка работы \cite{ms12} переобозначением константы $\lambda_1$ в уравнение \eqref{tst7}.

Высшие симметрии в дискретном направлении являются уравнениями типа:
\eq{u_{n,\tau}=G(u_{n-2},u_{n-1},u_{n},u_{n+1},u_{n+2}).\label{d_sym}} 



\section{Метод исследования.}

Из требования совместности уравнений \eqref{seq} и \eqref{scalar} получаем определяющее уравнение
\eq{V_{n+1,x}=\pfrac{f}{u_{n,x}}V_{n,x}+\pfrac{f}{u_{n+1}}V_{n+1}+\pfrac{f}{u_{n}}V_n, \label{lineq}}
где через $V_n$ обозначена правая часть уравнения \eqref{scalar}. Здесь используются обозначения:
\seq{V_{n,x}=\frac{d}{dx}V_n,\quad V_{n+1,x}=\frac{d}{dx}V_{n+1}}

Если функция $f$, определяющая правую часть уравнения \eqref{seq} известна, то из уравнения \eqref{lineq} можно находить правую часть высшей симметрии \eqref{scalar}. Эта процедура подробно описана в работе \cite{gh21}. Здесь же наоборот известна высшая симметрия, а само полудискретное уравнение не задано. Поэтому на функцию $f$ получается сложное нелинейное уравнение. Однако уравнение содержит дополнительные переменные, от которых не зависит функция $f$. Наличие этих переменных позволяет получать более простые дифференциальные следствия и определить неизвестную функцию $f$. 

\section{Результаты классификации.}
В данной секции приводятся найденные полудискретные уравнения и их высшие симметрии в дискретном направлении. Они сгруппированы по виду высшей симметрии в $x$ направлении. Верно следующее утверждение:
\begin{theorem} Если невырожденное нелинейное автономное уравнение \eqref{seq} допускает непрерывную высшую симметрию в виде одного из уравнений (\ref{tst1}--\ref{tst15m}), то оно имеет вид \seqref{tz} -- \seqref{hyp_4}.

\end{theorem}

{\bf Схема доказательства. } Правые части уравнений (\ref{tst1}--\ref{tst15m}) брались в качестве функции $V_n$ в определеяющем уравнении \eqref{lineq}. Для каждого из этих уравнений находились все возможные функции $f$ -- правые части уравнений \eqref{seq}. Только для четырех уравнений из списка (\ref{tst1}--\ref{tst15m}) получен положительный результат. Ниже приводится список найденных полудискретных уравнений (для них используется спициальная нумерация вида (S...), и их дискетные высшие симметрии. Разные уравнения списка (\ref{tst1}--\ref{tst15m}) разделяются с использованием символа ${{\bigstar}}$. 

\itemaa 
Для уравнения \eqref{tst7} полудискретное уравнение имеет вид: 
\mmeq{u_{n+1,x}=u_{n,x}+\lambda_1(e^{-2u_n}+e^{-2u_{n+1}})+\lambda_2\sqrt{e^{2u_n}+e^{2u_{n+1}}}\label{tz}}
Это уравнение было найдено в статье \cite{gh21}

\itemaa Для уравнения \eqref{tst8} найдено полудискретное уравнение:  
\mmeq{\label{dhypt1}
u_{n+1,x}=u_{n,x}+\lambda_1(e^{u_n}+e^{u_{n+1}})+\frac{\lambda_2}{\lambda_1}e^{-u_{n}-u_{n+1}}.}

\itemaa Для уравнения \eqref{tst13m} полудискретные уравнения имеют вид: 
\mmeq{f(u_{n+1,x})+u_{n+1,x}=(f(u_{n,x})+u_{n,x})A(u_{n+1},u_{n})\label{hyp_3},}
где функция $f(x)$ - решение кубического уравнения \eqref{alg1}, функция $A(u_{n+1},u_{n})$ определяется из алгебраического уравнения:
\eq{(A^3-1)^3c+27(w(u_n)-w(u_{n+1})A^2)^2(Aw(u_n)-w(u_{n+1}))A^2=0}
Наряду с представлением \eqref{hyp_3} можно выписать второе выражение:
\eq{(2f(u_{n+1,x})-u_{n+1,x})A^2(u_{n+1},u_{n})=2f(u_{n,x})-u_{n,x}\label{hyp_3_1}.} Система \seqref{hyp_3},\eqref{hyp_3_1} позволяет однозначно выражать $u_{n+1,x}$ или $u_{n,x}$ через остальные функции:
\seqs{3u_{n+1,x}=2(u_{n,x}+f(u_{n,x}))A(u_{n+1},u_{n})+(u_{n,x}-2f(u_{n,x}))A^{-2}(u_{n+1},u_{n}),\\ 
3u_{n,x}=2(u_{n+1,x}+f(u_{n+1,x}))A^{-1}(u_{n+1},u_{n})+(u_{n,x}-2f(u_{n,x}))A^{2}(u_{n+1},u_{n}).}

\itemaa Для уравнения \eqref{tst14m} полудискретные уравнения имеют такой же вид: 
\mmeq{f(u_{n+1,x})+u_{n+1,x}=(f(u_{n,x})+u_{n,x})B(u_{n+1},u_{n})\label{hyp_4}.}
 В этом случае функция $B(u_{n+1},u_{n})$ определяется из уравнения:
\eq{(B^6-1)^3c+27(B^2w(u_{n+1})-w(u_n))(B^4w(u_n)-w(u_{n+1}))^2B^4=0.} Здесь также есть второе представление:
\eq{(2f(u_{n+1,x})-u_{n+1,x})B^2(u_{n+1},u_{n})=2f(u_{n,x})-u_{n,x}\label{hyp_4_1}.}
 Система \seqref{hyp_4},\eqref{hyp_4_1} позволяет однозначно выражать $u_{n+1,x}$ или $u_{n,x}$ через остальные функции.

\begin{hyp}Все уравнения списка \seqref{tz} -- \seqref{hyp_4} обладают дискретными высшими симметрями вида \eqref{d_sym}.\end{hyp}
Для уравнений \seqref{tz} и \seqref{dhypt1} дискретные симметрии найдены. Для уравнения \seqref{tz} она имеет вид:
\eqa{\partial_\tau u=\Big((v^2-1)^2-4v_{-1}^2T^{-1}\Big)\frac{(v_{1}^2+1)(v_{-1}^2+1)}{(v^2(v_{-1}+1)^2+(v_{-1}-1)^2)(v_1(v+1)^2+(v-1)^2)},\\v=\sqrt{1+e^{2(u-u_1)}}+e^{u-u_1}.\label{u_tau}}
Для уравнения \seqref{dhypt1}:
\eqa{\label{voltI1}
\frac{du_{n}}{d\tau}=\frac{e^{u_{n+1}}e^{u_{n-1}}}{e^{u_{n+1}}+e^{u_n}+e^{u_{n-1}}}(T-T^{-1})\left(\frac{1}{e^{u_{n+1}}+e^{u_{n}}+e^{u_{n-1}}}-\frac{1}{e^{u_{n}}}\right).}
Для уравнений \eqref{hyp_3} и \eqref{hyp_4} поиск таких симметрий наталкивается на вычислительные трудности, но автор верит, что они существуют.

\section{Пары Лакса для уравнений \eqref{tst8}, \seqref{dhypt1}.}

В статье \cite{g21} были найдены представления Лакса для уравнений \eqref{tst7}, \seqref{tz}. Здесь с помощью аналогичных методов выпишием такие представления для второй пары уравнений \eqref{tst8}, \seqref{dhypt1}.

\subsection{Пара Лакса для уравнения \eqref{tst8}.}
Для решения этой задачи мы используем известную связь \cite{ms12} 
\eq{w=-u_{n,xx}-u_{n,x}^2+\lambda_1e^{u}u_{n,x}-\lambda_1^2e^{2u}+\lambda_2^2e^{-u}\label{wu}}между уравнением \eqref{tst8} и уравнением  Савады-Котерры \eqref{tst1}, см. \cite{sk74}.  Уравнение \eqref{tst1}, записанное в переменной $w$, имеет представление Лакса \cite{k80,fg80}\eq{\widetilde L_t=[\widetilde L,\widetilde A],\label{lax}} где \eqa{\widetilde L&=\partial^3+w\partial,\\ \widetilde A&=9(\widetilde L^{5/3})_+=9\partial^5+15w\partial^3+15w_x\partial^2+5(w^2+2w_2)\partial.\label{laxkk}} Здесь $\partial$ обозначает оператор дифференцирования по $x$, обозначение $()_+$ означает положительную часть формального ряда по степеням оператора $\partial.$

Если мы в представление Лакса \eqref{laxkk} подставим замену \eqref{wu}, то полученные операторы будут конечно же совместны на решениях \eqref{tst8}, но из их условия совместности \eqref{lax} будет следовать не уравнение \eqref{tst8}, а некоторое дифференциальное следствие этого уравнения. Для получения настоящего представления Лакса уравнения \eqref{tst8} нужно заметить что оператор $\widetilde L$ в переменной $u$ допускает следующую факторизацию:
\seq{\widetilde L=\left(\partial+u_{n,x}-\lambda_1e^{u}\right)\left(\partial^2-(u_{n,x}-\lambda_1e^{u})\partial+u_{0,2}+\lambda_2e^{-u}\right)+\lambda_1\lambda_2.}
Тогда в качестве оператора $L$ для уравнения \eqref{tst8} можно использовать оператор \eq{{L_1}=\left(\partial^2-(u_{n,x}-\lambda_1e^{u})\partial+u_{0,2}+\lambda_2e^{-u}\right)\left(\partial+u_{n,x}-\lambda_1e^{u}\right)+\lambda_1\lambda_2.}он несколько упрощается после преобразования: \eq{{L}=e^{u}{L_1}e^{-u}=\partial^3-3u_x\partial^2-(u_{n,xx}-2u_{n,x}^2+\lambda_1^2e^{2u}-\lambda_2e^{-u})\partial.\label{L_t}} 
Оператор $A$ приобретает вид:\eqa{{A}=&9\partial^5-45u_{n,x}\partial^4-15(4u_{n,2}-5u_{n,x}^2+\lambda_1^2e^{2u}-\lambda_2e^{-u})\partial^3\\-&45(u_{n,xxx}-3u_{n,xx}u_{n,x}+u_{n,x}^3+\lambda_2u_{n,x}e^{-u})\partial^2-5\big(2u_{n,xxxx}\\-&11u_{n,xxx}u_{n,x}+13u_{n,xx}u_{n,x}^2-6u_{n,xx}^2-u_{n,x}^4-(\lambda_1^2e^{2u}-\lambda_2e^{-u})^2\\+&\lambda_1^2e^{2u}(2u_{n,xx}+3u_{n,x}^2)+\lambda_2e^{-u}(2u_{n,xx}-3u_{n,x}^2)\big)\partial.\label{A_t}}

Верно следующее утверждение:
\begin{lemma} Уравнение \eqref{tst8} эквивалентно представлению Лакса \eq{L_t=[L,A] \label{op_t_eq},} где операторы $L$ и $A$ определены выражениями \eqref{L_t}  и \eqref{A_t}.
\end{lemma}

\subsection{Представление Лакса для уравнения \seqref{dhypt1}}
Уравнение \seqref{dhypt1} должно иметь представление Лакса \eq{(TL)M-ML=0,\label{MLo}} с некоторым оператором $M$, где $T$ -- оператор сдвига по дискретной переменной $n$. Оператор $M$ ищем в аналогичном виде, как в работе \cite{g21}:
\eqa{M&=M_3(u_{n+1},u_{n})\partial^3+M_2(u_{n+1},u_{n},u_{n,x})\partial^2\\&+M_1(u_{n+1},u_{n},u_{n,x},u_{n,xx})\partial^1+M_0(u_{n+1},u_{n},u_{n,x},u_{n,xx},u_{n,xxx}).}
Из требования выполнения операторного соотношения \eqref{MLo} на решениях уравнения \seqref{dhypt1} при разных степенях $\partial$ последовательно получаем уравнения. При $\partial^5$ имеем: \eqa{3&\left(\pdfrac{M_3}{u_{n+1}}+\pdfrac{M_3}{u_{n}}\right)u_{n,x}+\frac{3}\lambda_1\left(M_3-\pdfrac{M_3}{u_{n+1}}\right)(\lambda_1^2e^{u_{n+1}+u_{n}}+\lambda_2e^{-u_{n+1}-u_{n}})=0,\nonumber}
Решение которого: \eq{M_3=e^{u_{n+1}-u_{n}}\label{sol_M3}} Константу интегрирования взяли равной единицей, так как уравнение \eqref{MLo} однородно по оператору $M$. Остальные коэффициенты оператора $M$ находятся с точностью до констант из уравнений при $\partial^4,\partial^3,\partial^2$. Уравнения при $\partial^1,\partial^0$ определяют эти константы. Окончательно, оператор $M$ имеет вид:
\eqa{M=e^{u_{n+1}-u_{n}}(\partial^3-3u_{n,x}\partial^2-(u_{n,xx}-2u_{n,x}^2)\partial)\\+\frac{\lambda_2}{\lambda_1}e^{-2u_n}(\partial^2-u_{n,x}\partial)-\lambda_1^2e^{u_{n+1}+u_{n}}\partial-\lambda_1\lambda_2.\label{o_M}}
Оператор $M$ допускает факторизацию:
\seqs{M=e^{u_{n+1}-u_{n}}\left(\partial-2u_{n,x}+\frac{\lambda_2}{\lambda_1}e^{-u_{n+1}-u_{n}}+\lambda_1e^{u_{n+1}}\right)\\ \ast(\partial^2-(u_{n,x}+\lambda_1e^{u_{n+1}})\partial-\lambda_1^2(e^{2u_n}+e^{u_{n+1}+u_{n}}))}
Верно следующее утверждение:
\begin{lemma} Уравнение \seqref{dhypt1} эквивалентно представлению Лакса \eqref{MLo}, где операторы $L$ и $M$ определены выражениями \eqref{L_t}  и \eqref{o_M}.
\end{lemma}


\bigskip
\begin{thebibliography}{999}

\bibitem{g23}Р. Н. Гарифуллин, “Классификация полудискретных уравнений гиперболического типа. Случай симметрий третьего порядка”, ТМФ, 217:2 (2023), 404-415

\bibitem{vs93}А.П. Веселов, А.Б. Шабат, \emph{Одевающая цепочка и спектральная теория оператора Шрёдингера}, Функц. анализ и его прил., {\bf 27}:2, 1–21 (1993). 

\bibitem{g21}Р. Н. Гарифуллин, “Об интегрируемости полудискретного уравнения Цицейки”, Уфимск. матем. журн., 13:2 (2021), 18-24

\bibitem{y90} Р.И. Ямилов, {\emph Обратимые замены переменных, порожденные преобразованиями Беклунда}, Теор. и мат. физ. {\bf 85}:3, 368-375 (1990).

\bibitem{gh21}R.N.Garifullin and I.T.Habibullin \emph{Generalized symmetries and integrability conditions for hyperbolic type semi-discrete equations}, Journal of Physics A: Mathematical and Theoretical, {\bf 54}:20, 205201, 19 pp (2021).


\bibitem{ms12} А.Г. Мешков, В.В. Соколов, \emph{Интегрируемые эволюционные уравнения с постоянной сепарантой}, Уфимск. матем. журн., {\bf 4}:3,  104–154 (2012).  [Engl. trans.: Ufa Math. Journal {\bf 4}:3. 104--152 (2012).]

\bibitem{y06} R. Yamilov,  \emph{Symmetries as integrability criteria for differential difference equations}. Journal of Physics A: Mathematical and General, {\bf 39}(45), R541 (2006).

\bibitem{y83}Р.И. Ямилов, \emph{О классификации дискретных эволюционных уравнений}, Успехи мат. наук {\bf 38}:6, 155-156  (1983).

\bibitem{k80} D.J. Kaup. {\it On the Inverse Scattering Problem for Cubic Eigenvalue Problems of the Class $\psi_{xxx}+6Q\psi_x+6R\psi=\lambda\psi$} // Stud. Appl. Math. {\bf 62}, 189--216 (1980).

\bibitem{fg80}A.P. Fordy and J. Gibbons \emph{ Factorization of operators I. Miura transformations}, Journal of Mathematical Physics, 21(10), 2508--2510  (1980).

\bibitem{sk74}K. Sawada and  T. Kotera. {\it A method for finding {$N$}-soliton solutions of the {K}.d.{V}. equation and {K}.d.{V}.-like equation} // Progr. Theoret. Phys. {\bf 51.} 1355--1367 (1974).


\end{thebibliography}
\end{document}